\documentclass[amsmath,aps,twocolumn]{revtex4}
\usepackage{amsthm,amsfonts,graphicx,verbatim, color}

\newcommand{\Tr}{{\rm Tr}}
\newcommand{\tr}{{\rm tr}\/}

\newcommand{\D}{{\rm d}}

\newcommand{\be}{\begin{equation}}
\newcommand{\ee}{\end{equation}}
\newcommand{\bea}{\begin{eqnarray}}
\newcommand{\eea}{\end{eqnarray}}

\newcommand{\la}{\langle}
\newcommand{\ra}{\rangle}

\renewcommand{\epsilon}{\varepsilon}
\def\nn{\nonumber\\}

\begin{document}
\title{Birman-Schwinger and the number of Andreev states in BCS superconductors}
\author{Israel Klich}
\affiliation{ 
Department of Physics, University of Virginia,
Charlottesville VA 22904}

\begin{abstract}
The number of bound states resulting from inhomogeneities in a BCS superconductor is usually established either by variational means or via exact solutions of particularly simple, symmetric perturbations. Here, we propose estimating sub-gap states using the Birman-Schwinger principle.
We derive upper bounds on the number of sub-gap states for small normal regions and find a suitable Cwikel-Lieb-Rozenblum inequality. We also estimate the number of such states for large normal regions using high dimensional generalizations of the Szego theorem. The method works equally well for local inhomogeneities of the order parameter and for external potentials.
\end{abstract}
\maketitle
\section{Introduction}
It is well known that inhomogeneities in a superconducting system may give rise to Andreev bound states. Such bound states form in superconductor-normal metal-superconductor (SNS) junctions and can affect the transmission and noise properties of the junction.
The presence of such states in the cores of vortices is of much interest as well.  Vortices in s-wave superconductors may have Caroli-de-Gennes-Matricon (CdGM) states \cite{caroli1964bound}, bound states in vortices in $p+ip$ superfluids where studied e. g. in \cite{kopnin1991mutual}. Zero modes in the core of half-quantum vortices in p-wave superconductors \cite{volovik1976line,volovik1999monopole} are of extensive recent interest.  Such modes are described in terms of unpaired Majorana fermions and have non-abelian braiding statistics \cite{ivanov2001non,read2000paired}. Other inhomogeneities in the system have  been of interest as well, for example, stripes and other structures in the pairing function of a superconductor (See e.g. \cite{martin2005enhancement,zou2008effect}).

Often, the bound sates are either directly obtained for systems of high symmetry, such as a circular vortex, or idealized planar SNS junctions or studied using perturbation theory. In addition, multiple scattering and semi-classical (WKB) methods based on the Andreev approximation have been developed \cite{adagideli2002density}.

In this paper, we introduce an alternative view on Andreev states, complementing the above-mentioned methods. Our approach does not assume a particular symmetry, and works even when the Andreev approximation is inadequate  (see e.g. \cite{lesovik1998nonlinear}). 
We base our approach on a simple adaptation of the Birman-Schwinger counting argument to the superconducting scenario. 
The Birman-Schwinger method has been widely used to study bound states of the Shcr\"{o}dinger equation with a potential. 
The Birman-Schwinger bound has been 
introduced by Birman \cite{birman1961spectrum} and Schwinger \cite{schwinger1961bound} as a refinement of previous estimates by Bargmann \cite{bargman1952on} on the number of bound states in a potential.
For a discussion and review see \cite{reed1978analysis}.
Recently the Birman-Schwinger approach was used in the context of estimating the critical temperature of BCS models in \cite{frank2007critical,hainzl2008bcs,hainzl2008critical}.

In this paper we establish the following results: 

1. A bound on the number of sub-gap states in a superconductor containing a normal region. The bound on the number of states below a given sub-gap energy $E$  is expressed in eq. \eqref{BS for s wave}, and holds for general values of the chemical potential $\mu$, the gap $\Delta_{0}$. {  
For example,  in the limit of large chemical potential, the inequality is given in the simple form 
\begin{eqnarray}\label{BS leading mu1}N_{E}\leq
\frac{\sqrt{\mu } \Delta
   _0^4vol(A)}{\pi  \left(\Delta
   _0^2-E^2\right){}^{
   3/2}}+\frac{ \Delta
   _0^2 \left(\Delta _0^2-2
   E^2\right)vol(A)}{8 \pi   {
   \mu }^{3/2} 
\sqrt{\Delta_0^2-E^2}} \end{eqnarray} 
where $vol(A)$ is the volume of the normal region. Clearly, the inequality is most effective for energies not very close to the top of the gap. 

2. In studying the bound states of Schordinger operators with deep potential wells, often Birman-Schwinger inequalities yield poor upper bounds to the number of bound states. In such cases the Cwikel-Lieb-Rozenblum inequality \cite{cwikel1977weak, MR0295148, lieb1980number} may be used to obtain upper bounds with power low behavior in-line with the semi-classical expectations for deep wells. We establish the Cwikel-Lieb-Rozenblum inequality  \eqref{CLR_for_BdG} for the Bogolubov de Gennes equation. This bound requires more work and has the advantage of retaining a simple form for a spatially varying $\Delta$, however, for normal regions, and in the typical situation of gap energy being small compared to Fermi energy, it seems to give a worse upper-bound to the number of states than the simple Birman-Schwinger inequality.}

3. Finally, using semiclassical tools and the relation to Szeg\"o theory we supply an asymptotic expression for the number of states in large regions and a conjecture \eqref{conjecture} for the scaling $l^{{d-1}}\log(k_{F}l)$ of the sub-leading term. 

 {  The paper is organized as follows: In section \ref{sec:BS principle}, we derive a Birman-Schwinger operator which is suitable for the treatment of mean field BCS superconductors. In section \ref{sec:BS simple} we derive the basic bound on the number of states in a normal region of a superconductor. We proceed to present the Cwikel-Lieb-Rozenblum inequality in section \ref{sec:CLR}, the derivation of which is technical and deferred to the appendices. In section \ref{sec:Szego} we consider the boundary correction to the number of states for asymptotically large normal regions and explain their connection to the theory of Toeplitz operators and Szego theorems.  
}
\section{The Birman-Schwinger principle for BCS}\label{sec:BS principle}
To employ this method, we consider a superconductor described by the mean field BCS Hamiltonian:
\begin{eqnarray}&
H_{BCS}=\sum_{\sigma}\int \psi_{\sigma}^{\dag}(x)(u(x)-\mu-{\hbar\over 2m}\nabla^2)\psi_{\sigma}(x)\D x+\\ \nonumber & \int [\Delta_{\sigma \sigma'}(x,x')\psi_{\sigma}^{\dag}(x)\psi_{\sigma'}^{\dag}(x')+h.c.]\D x \D x'\, ,\end{eqnarray}
where $\Delta$ is the superconducting order parameter, $\mu$ is the chemical potential $u(x)$ is a local potential and $m$ the effective electron mass. $\Delta$ is related to the anomalous Green's function through a self consistency condition. Here we will assume that $u(x)$ and $\Delta$ are given, and represent the effective values of these parameters, which are, in principle, measurable directly.  

To diagonalize the BCS Hamiltonian we use the Bogolubov de Gennes (BdG) equation 
\begin{eqnarray}
H_{BdG}
\left(
\begin{array}{c}
\psi_e  \\
\psi_h
\end{array}
\right)=E\left(
\begin{array}{c}
\psi_e \\
\psi_h
\end{array}
\right)\,\,\,\,\,\,;\,\,\,H_{BdG}=
\left(
\begin{array}{cc}
h &  \Delta  \\
\Delta^*  &  -h
\end{array}
\right)
\end{eqnarray}
where $\psi_e,\psi_h$ are the electron and hole parts of the quasiparticle wave function, and $E$ is the quasi particle excitation energy measured relative to the Fermi energy. The solutions of this equation come in $+E,-E$ pairs (unpaired states may exist at $E=0$).
We will mostly consider $h=-\nabla^2-\mu+u(x)$ (Throughout, we take ${\hbar^{2}\over 2m}=1$).

For a translationally invariant $\Delta$ and $u(x)=0$, the spectrum of the BdG operator is continuous, and consists of the positive and negative energy bands. In an s-wave superconductor these bands are separated from zero by the gap $|\Delta|$. The method will also work for other gap functions, for example in a p-wave superconductor or superfluid with gap $\Delta(p)\sim \Delta_{0}(p_{x}+ip_{y})$ 
(as in the Anderson-Brinkman-Morel phase of $^{3}He$). Andreev states in d-wave superconductors have also been of much interest, however in in d-wave superconductors the gap function vanishes on the nodal lines and one needs to differentiate between resonant states and the rest of the spectrum.

To study the introduction of bound states into the system from an inhomogeneous $\Delta$ (and possibly also $u$), we write the perturbed $H_{BdG}$ as $H_{0}+W$ where $H_{0}$ is translationally invariant. Here $W$ represents the perturbation (inhomogenous $\Delta$, disorder potential etc.). We choose a decomposition of the perturbation $W$ into a product $W_1W_2$, and define the Birman-Schwinger operator:
\begin{eqnarray}\label{k operator}
K_E=W_2{1\over E-H_0}W_1
\end{eqnarray}
The key property of $K_E$ is that it has an eigenvalue ${1\over \lambda}$  for each eigenvalue $E$ of $(H_0+\lambda W)$. Indeed: 
\begin{align}&
(H_{0}+\lambda W)\phi=E\phi ~\Rightarrow~  {1\over \lambda}\phi=-(H_{0}-E)^{-1}W_{1}W_{2}\phi \\ \nonumber  &  \Rightarrow
 {1\over \lambda} (W_{2}\phi)= W_{2}(H_{0}-E)^{-1}W_{1}(W_{2}\phi)=K_{E}(W_{2}\phi)
\end{align}
Thus, $W_{2}\phi$ is an eigenvector of $K_E$ with eigenvalue ${1\over\lambda}$. The two major differences between the BCS system and the Birman-Schwinger kernel $K_E$ typically used to study bound states in Schrodinger operators is that $H_{BdG}$ is not bounded below, and that $K_E$ is not Hermitian, but the ideas have been used to estimate the number of states within a spectral gap (see e.g.\cite{deift1986existence}).

\begin{figure}[hb]
\includegraphics*[width=2in]{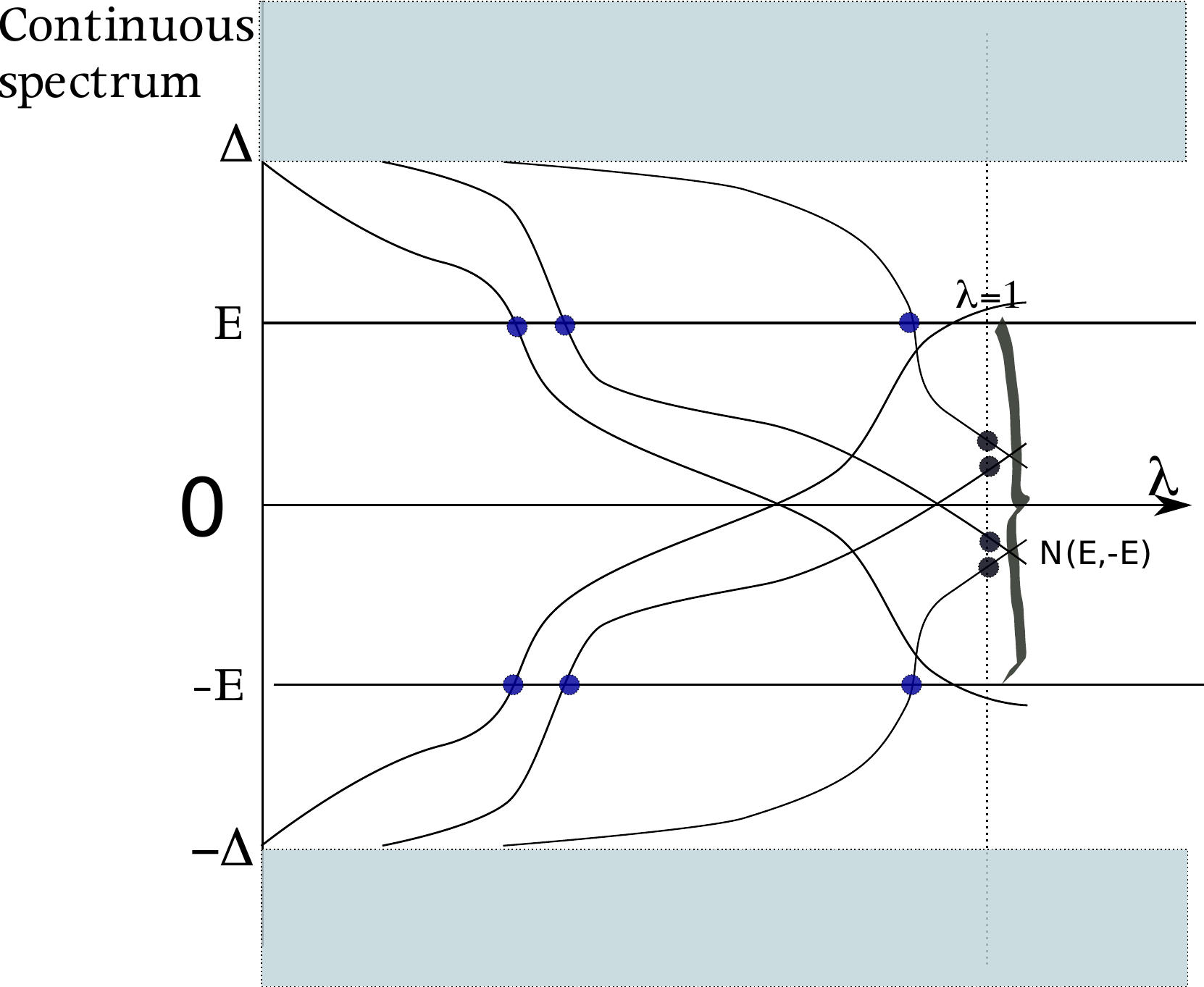}
\caption{The Birman-Schwinger principle: continuity of the bound state energy as function of $\lambda$ implies that each bound state at $\lambda=1$ with energy in $(E,-E)$ corresponds to at least one crossing of the $E$ or $-E$ line for some $\lambda\in(0,1)$.
} 
\label{fig1}
\end{figure}

We now invoke the Birman-Schwinger argument:

Let $N_{E_1,E_2}(V)=$ number of eigenvalues of $H_0+V$ in the interval $(E_1,E_2)$, where $E_{1},E_{2}$ are in the gap, i.e. $|E_{1}|,|E_{2}|<{\rm min}_p(|\Delta(p)|)$.
Assuming that the energy of eigenvalues of $H_{0}+\lambda V$ depends continuously on $\lambda$ as we increase $\lambda$ from $0$ to $1$, we see that each eigenvalue for $\lambda=1$ must have crossed the $E_1$ or $E_2$ line for some value of $\lambda\in(0,1)$, as illustrated in Fig. \ref{fig1}. Thus:{
\begin{eqnarray}&
\label{BS principle}N(E)\leq ~number~of~crossings~fot~\lambda\in(0,1)=\nn &\# eigenvalues~of~K_{E}~larger~than~1
\end{eqnarray}}


Note that the inequality sign in \eqref{BS principle} is due to the fact that in all cases we consider perturbations that are not strictly positive or negative. The curves in Fig. \ref{fig1} may be non monotonous functions of $\lambda$ and an in-gap eigenvalue at $\lambda=1$ may correspond to several crossings. Moreover, in  some situations no bound states will be created even for very large $\lambda$. 

It is important to note that for unpaired zero modes of an unperturbed system such as the $p_{x}+ip_{y}$ superconductor (for example in the presence of vortex or boundary), the principle will still work. Such a zero mode is stable as long as it is far from other vortices, and will not shift when adding impurity potentials. 

We now find the explicit form of the Birman-Schwinger kernel useful for the study of Andreev bound states in a normal domain of general shape embedded in the superconductor. First, we decompose: $\Delta=\Delta_0+\delta\Delta$,  $\delta\Delta=\delta\Delta_1\delta\Delta_2$, and note that \begin{eqnarray}
\left(
\begin{array}{cc}
0 &   \delta\Delta  \\
\delta\Delta^* & 0
\end{array}
\right)
=\left(
\begin{array}{cc}
\delta\Delta_1 &   0  \\
0  & \delta\Delta_2^*
\end{array}
\right)
\left(
\begin{array}{cc}
0 &   \delta\Delta_2  \\
\delta\Delta_1^*  & 0
\end{array}
\right)
\end{eqnarray}

\begin{figure}[hb]
\includegraphics*[width=3in]{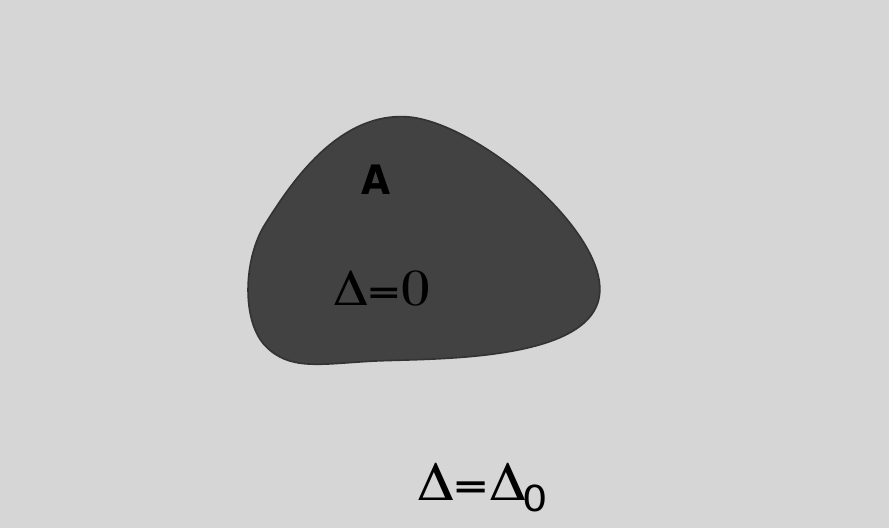}
\caption{A normal region $A$ inside the superconductor.
} 
\label{fig2}
\end{figure}

To further simplify the system we assume an $s$ wave superconductor. This will be described by $\Delta=\Delta_0$ outside the region and $\Delta=0$ inside (Fig. \ref{fig2}). Thus we write $\delta\Delta_{1}=\sigma_{x}\chi_{A}(x)$ and $\delta\Delta_{2}=-\Delta_{0}{\bf I}\chi_{A}(x)$, where ${\bf I}$ is the identity matrix,
and $\chi_{A}(x)$ is the characteristic function of the region $A$, so $\chi_{A}(x)=1$ if $x\in A$ and $0$ otherwise.
The Birman-Schwinger operator \eqref{k operator} can be written explicitly. Assuming real $\Delta_{0}$ for simplicity, we find:
\begin{eqnarray}\label{Birman Schwinger kernel for s wave}&
K_{A,E}(x,x')=\nn & {\Delta_{0}\over \sqrt{{\Delta_{0}}^2-E^2}}(\sigma_{x} E-\Delta_{0})\chi_{A}(x)\rm{Im}\, g_{+}(x-x')\chi_{A}(x') \nn & -i{\Delta_{0}}\sigma_y\chi_{A}(x)\rm{Re}\, g_{+}(x-x') \chi_{A}(x')
\end{eqnarray}
where
\begin{eqnarray}
g_{\pm}(x-x')={e^{\pm ik_{\pm}|x-x'|}\over 4\pi|x-x'|}
\end{eqnarray}
where $k_{\pm}=\sqrt{\mu\pm i\sqrt{{\Delta_{0}}^2-E^2}}$, and ${\rm Re}\,k>0$.

Note that from the numerical point of view studying the Birman-Schwinger kernel \eqref{Birman Schwinger kernel for s wave} has an immediate advantage: the kernel is only supported on the region $A$. Thus, computation of the spectrum of bound states requires merely finding the eigenvalues of this operator restricted to the region $A$. Since the operator in \eqref{Birman Schwinger kernel for s wave} is known analytically, one can numerically compute eigenvalues by simply discretizing the region $A$.

Let us proceed and consider limits where the application of these ideas is particularly interesting:
\section{normal regions:}\label{sec:BS simple} 
Here we use $K_{{A,E}}$ to obtain an upper bound to the number of states. 
The basic inequality is:
\begin{eqnarray}\label{Birman Schwinger inequality} &
N_E(V)\leq \sum_{\lambda|\mu_n(\lambda)=E_1\,{\rm \,\,or\,\, }E_2 \,{\rm for\,some\,}\lambda\in(0,1)} 1\leq \\ \nonumber  & \sum_{\lambda|\mu_n(\lambda)=E_1\,{\rm \,\,or\,\, }E_2 \,{\rm for\,some\,}\lambda\in(0,1)} {1\over \lambda^2}\leq \\ \nonumber   & \sum_{\lambda{\rm\,eigenvalue\,of}\,K_{E1}}{1\over \lambda^2}+\sum_{\lambda{\rm\,eigenvalue\,of}\,K_{E2}}{1\over \lambda^2} \leq \\ \nonumber  &\Tr {K}^{*}_{E1}K_{E1}+ \Tr K^{*}_{E2}K_{E2}
\end{eqnarray}
The last inequality is due to the fact that $K_E$, is, as opposed to the usual Birman-Schwinger case, not self-adjoint, therefore we used that for a compact operator $\sum_{eigenvalues} |\lambda|^p\leq\sum_{singular\,values} |s|^p$ for $p\geq 1$ (Schur-Lalesco-Weyl theorem).

This allows us to write a simple inequality for the number of sub-gap states. Plugging $K_{E,A}$ from \eqref{Birman Schwinger kernel for s wave} in \eqref{Birman Schwinger inequality} we find:
\begin{eqnarray}\label{BS for s wave}&
N(E)\leq 4\Delta_{0}^{2}\int_{A\times A}\D x\D x'\Big(
{\Delta_{0}^{2}+E^{2}\over {\Delta_{0}}^2-E^2}|\rm{Im}\, g_{-}(x-x')|^{2}+\nn &|\rm{Re}\, g_{+}(x-x')|^{2}\Big)
\end{eqnarray}

As $E\rightarrow 0$ the number of states $N_{V}(E)$ goes to zero, thus, the  estimate \eqref{BS for s wave} is effective only when the expression on the right hand side of \eqref{BS for s wave} is less than 1 in that limit. Let us estimate the size of the region $A$ where this holds, assuming a typical gap much smaller than $k_F$. Using that for $E\rightarrow 0$ ,  $k_{{\pm}}\sim \sqrt{\mu}\pm i{\Delta_{0}\over \sqrt{\mu}}$, we see that 
\begin{eqnarray}&
N(E)\leq 
 {\Delta_{0}^{2}\over 4\pi^{2}}\int_{A\times A}\D x\D x'{e^{-\sqrt{{\Delta_{0}^2-E^2\over \mu}}|x-x'|}\over |x-x'|^{2}}\times\nn &\Big(
{2E^{2}\over {\Delta_{0}}^2-E^2}\sin^{2}( \sqrt{\mu}|x-x'|)+1\Big)
\end{eqnarray}
and so the decay is set by coherence length  $\xi={\sqrt{\mu}\over \Delta{0}}$. 
This means that we need the condition ${1\over \pi}\sqrt{\mu}\Delta_{0}Vol(A)<1$ to make \eqref{BS for s wave} effective at low energies.

Much like for the Schrodinger equation in 3d, we see that for a small enough normal region, the right hand side of \eqref{BS for s wave} will be less than one, and so no sub-gap states will be present below this energy $E$.

Numerically, we have checked the utility of this approximation for simple situations on a lattice compared to the exact result. The agreement is fairly good for $E$ small $E<0.25\Delta_{0}$.

{Let us estimate the leading volume dependence of the bound \eqref{BS for s wave}. Writing  \begin{eqnarray}
g_{\pm}(r)={e^{\pm ik_{\pm}|r|}\over 4\pi|r|}={e^{-a r\pm ib |r|}\over 4\pi|r|}
\end{eqnarray}
with: 
\begin{eqnarray}&
a=(\mu^{2}+\Delta_{0}^{2}-E^{2})^{1/4}\sin[{1\over 2}\arctan{\sqrt{\Delta_{0}^{2}-E^{2}}\over \mu}] \nn &
b=(\mu^{2}+\Delta_{0}^{2}-E^{2})^{1/4}\cos[{1\over 2}\arctan{\sqrt{\Delta_{0}^{2}-E^{2}}\over \mu}]
\end{eqnarray}
By \eqref{BS for s wave}
we have: $N(E)\leq N_{1}$ where
\begin{eqnarray} &
N_1= 4\Delta_{0}^{2}\int_{A\times A}\D x\D x'\Big(
{\Delta_{0}^{2}+E^{2}\over {\Delta_{0}}^2-E^2}|\rm{Im}\, g_{-}(x-x')|^{2}+\nn & |\rm{Re}\, g_{+}(x-x')|^{2}\Big)
\end{eqnarray}
now, we change to variables $x,x'\rightarrow x,r=(x'-x)$. Taking the $r$ integral over all space we get that $N_1\leq N_{1}'$ with:
\begin{eqnarray*} &
N_{1}'=\nn &16\pi \Delta_{0}^{2}vol(A) 
\int_0^{\infty}r^{2}\D  r \Big(
{\Delta_{0}^{2}+E^{2}\over {\Delta_{0}}^2-E^2}|\rm{Im}\, g_{-}(r)|^{2}+|\rm{Re}\, g_{+}(r)|^{2}\Big)\nn & =
\Delta_{0}^{2} vol(A)
\int_0^{\infty} \D  r {e^{-2a r}\over \pi}\Big(
{\Delta_{0}^{2}+E^{2}\over {\Delta_{0}}^2-E^2}\sin^{2}(br)+\cos^{2}(br)\Big)
\end{eqnarray*}
This last integral can be carried out analytically yielding:
\begin{eqnarray} \label{BS leading}&
N_{1}'=
{\Delta_{0}^{2}\over 4\pi a} vol(A)
({1\over 2}+
 {{2E^{2}\over {\Delta_{0}}^2-E^2}\over \sqrt{\mu^{2}+\Delta_{0}^{2}-E^{2}}}
 b^2)
\end{eqnarray}
Finally, we can write $N_{1}'$ as:
\begin{eqnarray}\label{N1prime} &
N_{1}'=\frac{vol(A)\Delta _0^2
   \left((2 \Delta
   _0^2-E^2)
   \sqrt{\mu
   ^2+\Delta _0^2-E^2}+\mu E^2
    \right)}
   {4 \pi 
   \left(\Delta
   _0^2-E^2\right)
   \left(\mu
   ^2+\Delta
   _0^2-E^2\right)^{3\over 4} \sin
   \left(\frac{1}{2} {\tan}^{-1}
\frac{\sqrt{\Delta _0^2-E^2}}{\mu
   }
   \right)}
\end{eqnarray} 

The upper estimate $N_{1}'$ in \eqref{BS leading} is in fact a good apprximation to $N_{1}$ whenever the size of $A$ is larger than $1/a$ the decay scale for the Green's function.

The most physically common situation for BCS is $\mu>>\Delta_{0}$. In this limit we have:
\begin{eqnarray}\label{BS leading mu}&N_{1}'\sim
\frac{\sqrt{\mu } \Delta
   _0^4vol(A)}{\pi  \left(\Delta
   _0^2-E^2\right){}^{
   3/2}}+\frac{ \Delta
   _0^2 \left(\Delta _0^2-2
   E^2\right)vol(A)}{8 \pi   {
   \mu }^{3/2} 
\sqrt{\Delta_0^2-E^2}} \end{eqnarray} 
Taking $\mu$ fixed and $\Delta_{0}$ large, while keeping $\alpha=\Delta_{0}/E$ fixed, we get
\begin{eqnarray}\label{BS leading delta}&
N_{1}'\sim\frac{\left(2-\alpha ^2\right
   ) \Delta _0^{3/2}}{2 \sqrt{2} \pi 
   \left(1-\alpha ^2\right)^{5/4}}+\frac{\left(\alpha ^2+2\right) \mu  \sqrt{\Delta _0}}{4
   \sqrt{2} \pi 
   \left(1-\alpha ^2\right)^{7/4}}\end{eqnarray}

 }

\section{Cwikel-Lieb-Rozenblum ineuqality}\label{sec:CLR} 
Often, the estimate \eqref{Birman Schwinger inequality} may be quite poor because of large contributions from small eigenvalues of $K_{{E,A}}$. This behavior may sometimes be remedied by considering other inequalities, such as the celebrated Cwikel-Lieb-Rozenblum inequality \cite{cwikel1977weak, MR0295148, lieb1980number}.

To do so we consider the BdG equation with a gap function $\Delta_{0}+\Delta_1(x)$. Using the ideas in \cite{cwikel1977weak,cancelier1996remarks} and adapting to the superconducting scenario we find that the inequality
\begin{multline}\label{CLR_for_BdG}
N(0,E) \leq C{\mu^{1/2}|\Delta_{0}|^{2}\over (|\Delta_{0}|^{2}-E^{2})^{2}} \\ \int_{|\Delta_1(x)|\geq{ |\Delta_{0}|^{2}-E^{2}\over 4|\Delta_{0}|}} \D x \Big[\Delta_{0}^{3/2}|\Delta_1(x)|^{3/2}+{|\Delta_1(x)|^{3}}\Big].
\end{multline}
is valid for $\mu >\frac{\sqrt{\Delta _0^2-E^2}}{18 \sqrt{2}}$, with $C$ a constant of order one. The details of this calculation are given in the appendices.

For the case $\mu <\frac{\sqrt{\Delta _0^2-E^2}}{18 \sqrt{2}}$, the inequality is changed to 
\begin{multline}\label{CLR_for_BdG case2}
N(0,E) \leq \frac{2^{-3/4}C}{3 (\Delta _0^2 -E^{2} )^{3/4}} \\ \int_{|\Delta_1(x)|\geq{ |\Delta_{0}|^{2}-E^{2}\over 4|\Delta_{0}|}} \D x \Big[\Delta_{0}^{3/2}|\Delta_1(x)|^{3/2}+{|\Delta_1(x)|^{3}}\Big].
\end{multline}

The bounds \eqref{CLR_for_BdG} and \eqref{CLR_for_BdG case2} are closely related to the bound of \cite{cancelier1996remarks} on the number of bound states of perturbed Dirac operators.
The main technical difference with \cite{cancelier1996remarks} is the presence of a chemical potential and thus a presence of a non vanishing Fermi surface   \footnote{Technically, these change the weak norms appearing in the Cwikel inequalities for singular numbers}.

Note that the denominator $(|\Delta_{0}|^{2}-E^{2})^{2}$ in \eqref{CLR_for_BdG} shows that this bound is only useful for $E< |\Delta_{0}|$. 
However, most of our interest here is actually with states that are not too close to the top of the gap itself. The reason for this is that in practice, if the BCS self consistency is taken into account, the first states that are affected are those with energy very close to the gap. Such states will often be washed out from the spectrum, since the gap function may decrease slightly in the vicinity of the normal region.

{
In the case of a system defined by a normal region we $\Delta_1(x)=-\Delta_{0}\chi_{A}(x)$, we have $N(E)\leq N_{CLR}$ with :
\begin{eqnarray}\label{CLR_for_BdGsimple}
N_{CLR}= C{2\mu^{1/2}|\Delta_{0}|^{5}\over (|\Delta_{0}|^{2}-E^{2})^{2}}vol(A)
\end{eqnarray}
Let us compare the simple estimates CLR and BS estimates.
Consider the large $\mu$ limit 
 \eqref{CLR_for_BdGsimple}  vs \eqref{BS leading mu}. 
 We have:
 \begin{eqnarray}
N_{1}'/N_{C}\sim {\sqrt{\Delta_{0}^{2}-E^{2}}\over \Delta_{0}}<1
\end{eqnarray}

We conclude that for the typical BCS situation, the simple Birman-Schiwnger bound \eqref{BS leading mu} is more effective in dealing with normal regions.
}

\section{Asymptotic regime, Szeg\"o theorem}\label{sec:Szego} 
Here we are interested in the large $l$ behavior of the eigenvalues of $K_{{E,lA}}$, where $lA$ is the region $A$  rescaled by a factor $l$ (i.e. $lA=\{{\bf x}: l^{-1} {{\bf x}}\in A\}$).
In this limit semi-classical methods are often useful. 
To study the eigenvalues of 
$K_{E,lA}$, it is convenient to characterize the behavior of $\Tr f(K_{E,lA})$ for various functions $f$. We immediately see, since in our situation $[H_{0}(p),\Delta]$ is supported only on the boundary of the system, that in the volume term we may commute the $\chi_{A}$ terms in the perturbations $W_{i}$ and the free propagator. This gives the asymptotic leading term for the number of sub-gap states as the number of eigenvalues of $K_{{E, la}}$ larger than one. The semiclassical approximation expresses it as an integral over the classical phase space:
\begin{eqnarray}
N(E,0)\sim l^{d}vol(A)\int \Theta[(H_{0}(p)-E)^{-1}\sigma_{x})-1]\D^{d} p
\end{eqnarray}
where $\Theta$ is the Heaviside step function.

We may view the operator $K_{E,A}$ as a higher dimensional block of a Toeplitz operator: the kernel depends only on $x-x'$, however the $x,x'$ indices are restricted to be in $A$. This is equivalent to chopping a ``block'' from a regular matrix. The theory describing the asymptotic properties of blocks of Toeplitz matrices is based on various Szeg\"o theorems, and appears in numerous problems in physics. Most famously,  the strong  (two-term) Szeg\"o limit theorem
was initially used in the celebrated computation of the spontaneous magnetization for
the 2D Ising model by Onsager (see e.g.~\cite{bottcher1995onsager}).
The Szeg\"o limit theorem also plays a
special role in entanglement entropy studies, see in particular \cite{jin2004quantum,gioev2006entanglement}.

To try and describe the next term, we turn to a higher dimensional generalization of the Szeg\"o limit theorem, which has been extensively studied by Widom \cite{widom1980szego}. 
Denote ${\cal G}_E(p)=(H_{0}(p)-E)^{-1}\sigma_{x}$.
Given a unit vector direction $n_x$, we can define an operator acting on functions of a single variable, with kernel :
\begin{eqnarray}{\cal G}_{E,n_x,p_{\perp}}(s-t)=\int e^{ip_{||}(s-t)}{\cal G}_E(p_{\perp}+n_x p_{||}) \D p_{||}.\end{eqnarray}
Using  \cite{widom1980szego} we see that 
\begin{eqnarray}&\label{Widom_szego}
\Tr f(K_{E,lA})\sim {l^{d}vol(A)\over (2\pi)^d}\int f[{\cal G}_E(p)]\D^{d} p+\nn &
{l^{d-1} \over (2\pi)^{d-1}}\int_{\partial A}\D x \int_{ p_{\perp}\cdot n_x=0}\D^{d-1} p_{\perp}  \tr \Big[f(\chi_{+}{\cal G}_{E,n_x,p_{\perp}}\chi_{+})\nn &-\chi_{+}f({\cal G}_{E,n_x,p_{\perp}})\chi_{+}\Big]+o(l^{d-1})
\end{eqnarray}
where $n_x$ is the normal to $\partial A$ at $x$.
For the case of an $s$-wave superconductor the operator ${\cal G}_{E,n_x,p_{\perp}}$ can be computed explicitly:
\begin{eqnarray}&
{\cal G}_{E,n_x,p_{\perp}}(s-t)=\nn & -\sqrt{\pi \Delta_0^2\over 2(\Delta_0^2-E^2)} {\rm Im} ({e^{-|s-t|k_{+}(p_{\perp})}\over k_{+}(p_{\perp})})(\Delta_0+z\sigma_x)\nn & +i\sqrt{\pi\over 2}{\Delta_0 }{\rm Re}( {e^{-|s-t|k_{+}(p_{\perp})}\over k_{+}(p_{\perp})} )\sigma_y
\end{eqnarray} 
where where $k_{\pm}(p_{\perp})=\sqrt{p_{\perp}^2-\mu\pm i\sqrt{{\Delta_{0}}^2-E^2}}$, and ${\rm Re}\,k_{\pm}>0$
(here $i\sigma_y=(0,1,-1,0)$).

In our case, we are interested in the number of eigenvalues above 1. Thus we need to solve this problem for $f(\lambda)=\Theta(\lambda-1)$. However, for this function the arguments leading to \eqref{Widom_szego} are no longer applicable, but rather, due to the discontinuous nature of $f$, the area term as written in \eqref{Widom_szego} is divergent. We conjecture that in such a case, the boundary term should be modified to:
\begin{eqnarray}\label{conjecture}&
{l^{d-1} \over (2\pi)^{d-1}}\int_{\partial A}\D x \int_{ p_{\perp}\cdot n_x=0}\D^{d-1} p_{\perp}  \tr \Big[f(\chi_{l+}{\cal G}_{E,n_x,p_{\perp}}\chi_{l+})\nn &-\chi_{l+}f({\cal G}_{E,n_x,p_{\perp}})\chi_{l+}\Big]+o(l^{d-1}),
\end{eqnarray}
where $\chi_{l+}(s)=\theta(0<s<l)$. In analogy to the appearance of logarithmic corrections when considering similar expressions for the entanglement entropy of fermions in d-dimensions, we expect this term to scale as $O(l^{{d-1}}\log l)$, which also seems to be consistent with a preliminary numerical investigation.

In this case, (as well as for any isotropic superconductor), this integral is independent of the direction $n_x$, and so we have that the boundary term is given by:
\begin{eqnarray}
{l^{d-1}\log l \over (2\pi)^{d-1}}vol({\partial A})m_f(E)
\end{eqnarray}
for some function $m_f(E)$. The transformation $f\rightarrow m_f$ is independent of the shape, and we believe can be studied using Wienner Hopf methods for specific cases.

Finally, we remark that it has been argued that there is a mini-gap present in the local density of states if the shape of the normal region corresponds to a chaotic ``Andreev billiard'' \cite{atland1996random,melsen1997superconductor,lodder1998density,Schomerus1999excitation}. The mini-gap is of order $\Delta_{0}\log(k_{F}l)$, and it's nature depends on $\tau_{E}/\tau_{D}$, where  $\tau_{D}$ is dwell time in the normal region and $\tau_{E}$ is the Ehrenfest time. This question has recently been revisited in \cite{kuipers2010semiclassical}. To arrive at this result, the strategy is to consider the BdG equation within the Andreev approximation and analyzing possible classical trajectories. It would be of interest to see whether such a result may come out of a Birman-Schwinger like analysis.

\section{Conclusions and outlook}
To summarize, in this paper, we studied the bound states in superconductors using a Birman-Schwinger approach. The advantage of the method is its validity for general perturbations, and beyond Andreev approximation. Our main results are stated in following equations: The bound on sub-gap states \eqref{BS for s wave}, the Cwikel-Lieb-Rozenblum type inequality for BdG \eqref{CLR_for_BdG} and the asymptotic expression for the number of states in large regions with the conjecture \eqref{conjecture} for the scaling $l^{{d-1}}\log(k_{F}l)$ of the sub-leading term. For most physical situations the Birman-Schiwnger bound \eqref{BS for s wave} seems a better bound than the CLR bound \eqref{CLR_for_BdG}, since it gives a lower upper bound for large fermi energy $\mu$. 

In the same way it is possible to study the effect of adding disorder potentials, as well as spatially varying or momentum dependent $\Delta$. Because of the wide interest in the properties of vortices in such systems it is also of great interest to understand if the Birman-Schwinger  methods can be useful in the study of  bound states in vortices. {Indeed, it is not immediately clear how this should be done: 
The presence of core states in vortices seems to be mainly due to the phase winding around the vortex core (see, e.g. \cite{Berthod05}), and arguably less sensitive to the supression of order parameter in the core. Since the phase winding is of a topological nature (i.e. it is discrete) it cannot be considered as a perturbation which is turned on in a continuous way from zero, as in the case we have considered. The present approach may be utilized to understand the interplay between the topological and non-topological state binding. To do so, one should take as the unperturbed vortex an idealized radially symmetric vortex, and then add phase gradients and order parameter suppression as perturbations, and study the additional bound states introduced in this manner.}
\\ \\
{\bf Acknowledgments}: I would like to thank A. Auerbach and A. Elgart for useful discussions, and the hospitality of the Aspen Center for Physics where this research was initiated. I acknowledge financial support from NSF grant No. DMR-0956053.
\\ \\

\section*{Appendix: Cwikel-Lieb-Rozenblum for BdG}
\subsection{Estimates on singular values}
To control the spectrum of the Birman Schwinger operators, 
we will use the following inequality ($p>2$), (Thm XI.22 in \cite{reed1979methods}):
\begin{eqnarray}\label{Cwikel inequality}
s_{n}(\hat{ab})\leq n^{-1/p}C_{p,d}||a||_{p}||b||_{p,w},
\end{eqnarray}
where $s_{n}$ is the $n$-th singular value of the operator $\hat{ab}=a(x)b(i\nabla)$, i.e. the $n$th largest eignvalue of the operator $\sqrt{(\hat{ab})^{{\dag}}\hat{ab}}$. Here $C_{p,d}$ is a constant, $||a||_{p}=(\int \D x |a(x)|^{p})^{1/p}$ is the $L^{p}$ norm, 
and the value $||b||_{p,w}$ is defined as follows (pg. 30 in \cite{reed1975methods}). Defining:
\begin{eqnarray}
M_{b}(t)=vol(x||b(x)|>t)
\end{eqnarray}
$||b||_{p,w}$ is given by:
\begin{eqnarray}\label{weak norm}
||b||_{p,w}=\inf_{C}\{C:M_{b}(t)<{C^{p}\over t^{p}} \forall t\}
\end{eqnarray}
Note $||b||_{p,w}$ is not strictly speaking a norm, since it does not satisfy a triangle inequality. It is called weak since $L^{p}\subset L^{w}_{p}$.

To familiarize ourselves with computing $||b||_{p,w}$ consider first the following example:

{\it Example:}  Let us compute $||b||_{3,w}$  for $b(x)=(x^{2}+c)^{-1/2}$ assuming $c>0$. We have: \begin{eqnarray}&
M_{b}(t)=vol(x|(x^{2}+c)^{-1/2}>t)=\nn &vol(x^{2}<{1\over t^{2}}-c)\theta({1\over t^{2}}-c>0)
\end{eqnarray}
In $3d$ this gives:
\begin{eqnarray}
M_{b}(t)={4\pi\over 3}|{1\over t^{2}}-c|^{3/2}\theta({1\over t^{2}}-c>0)\leq {4\pi\over 3}{1\over t^{3}}
\end{eqnarray}
since the above inequality approaches equality for $t\rightarrow 0$, the infimum in \eqref{weak norm}, is given by : $||b||_{3,w}=({4\pi\over 3})^{1/3}$. Note that $||b||_{3,w}$  does not depend on $c$.

In the derivation of the Cwikel-Lieb-Rozenblum inequality we will need the $||b||_{3,w}$ estimate for 
\begin{eqnarray} \label{b bdg}
b(i\nabla)=((-\nabla^{2}-m)^{2}+c)^{-1/2}, \end{eqnarray} with $m,c>0$. 
Here we show that \begin{eqnarray} 
||b||_{3,w}\leq c^{-{1\over 4}}({v_3 })^{1/3}max\Big[1, \big({324m^{2}\over{c}}\big)^{1/12} \Big]
\end{eqnarray}
As in the example above, we define 
\begin{eqnarray}&
M_{b}(t)=vol(x|((x^{2}-m)^{2}+c)^{-1/2}>t)\nn & =vol((x^{2}-m)^{2}<{1\over t^{2}}-c)\theta({1\over t^{2}}-c>0)
\end{eqnarray}
Let $v_d={\pi^{d/2}\over \Gamma(d/2+1)}$ be the volume of a $d$ dimensional sphere. Then:
\begin{multline}
M_{b}(t)=\\   \begin{cases} v_d(m+\sqrt{{1\over t^{2}}-c})^{d/2}~~~~{\rm if} ~~~~~0<t<(m^{2}+c)^{{-1/2}}   &  \\ 
v_d\Big[(m+\sqrt{{1\over t^{2}}-c})^{d/2}-(m-\sqrt{{1\over t^{2}}-c})^{d/2}\Big] ~~~~{\rm if} & \\ ~~~~~~~~~~~~~~~~~~~~~~~~~~~~~~~~~~~~~~(m^{2}+c)^{{-1/2}}<t<{1\over \sqrt{c}}  \\ 0~~~~~~~~~~~{\rm if}~~~~~~~~~  t>{1\over \sqrt{c}} &  \end{cases}
\end{multline}
From the small $t$ behavior, we see that to bound $M_{b}$ uniformly as in the definition \eqref{weak norm} the power has to be at least $1/t^{d/2}$, i.e. ($p\geq d/2$). In the first sector we see also that the constant has to be at least $v_d$. However $v_d$ may be not good enough since the bound in the region 2 may be different. Let us estimate it:

For $(m^{2}+c)^{{-1/2}}<t<{1\over \sqrt{c}}$,
\begin{eqnarray*} &
v_d\Big[(m+\sqrt{{1\over t^{2}}-c})^{d/2}-(m-\sqrt{{1\over t^{2}}-c})^{d/2}\Big] =\nn &
v_d\int_{-1}^{1}\D s \partial_{s}(m+s\sqrt{{1\over t^{2}}-c})^{d/2}= \nn &
{dv_d\over 2}\sqrt{{1\over t^{2}}-c}\int_{-1}^{1}\D s (m+s\sqrt{{1\over t^{2}}-c})^{d/2-1}\leq  {dv_d\over t} (2m)^{d/2-1}
\end{eqnarray*}
where we used that in this region $m\geq \sqrt{{1\over t^{2}}-c}$.

Now we have to write this as bounded by a $1/t^{d/2}$ expression. If $d/2<1$ we are done, but if not, $d/2-1>0$, like $d=3$ case, using ${1\over t\sqrt{c}}>1$ we have:
\begin{eqnarray}&
{dv_d\over t} (2m)^{d/2-1}\leq {dv_d\over t} (2m)^{d/2-1} {1\over t^{d/2-1}(\sqrt{c})^{d/2-1}}=\nn &{dv_d  (2m)^{d/2-1} \over(\sqrt{c})^{d/2-1} t^{d/2}}
\end{eqnarray}
Combining all these results we see that (for $d>2$)
$$M_{b}(t)\leq max(1, {d (2m)^{d/2-1} \over(\sqrt{c})^{d/2-1}}){v_d \over t^{3/2}}$$
And thus we see that in this case we are in the weak $L^{3/2}_{w}$ space. 
i.e.: $||b||_{3/2,w}\leq({v_d }max(1, {3 (2m)^{1/2} \over{c}^{1/4}}))^{2/3}$.

The Cwikel estimates  require $p>2$. Let us estimate $||b||_{3,w}$. Since $t^{-3/2}<t^{-3}$, say, for $t<1$, we need to deal with the large $t$ behavior of $M_b$. Let us use the following:\\
\underline{{\it Lemma:}}
If $M_{b}<C t^{-\alpha}$, and $M_{b}=0$ for $t>t_{0}$, then we have, for any $\alpha_{1}>\alpha$
that: \begin{eqnarray} ||b||_{\alpha_{1},w}\leq (t_{0}^{\alpha_{1}-\alpha}C)^{1/p} .\end{eqnarray}
{\it Proof:} for $t<t_{0}$ we have $$M_{b}(t)\leq Ct^{-\alpha}<Ct^{-\alpha} (t_{0}/t)^{\alpha_{1}-\alpha} <Ct^{-\alpha_{1}} t_{0}^{\alpha_{1}-\alpha},$$ and $M_{b}=0$ for $t>t_{0}$ therefore $M_{b}<Ct_{0}^{\alpha-\alpha_{1}} t^{-\alpha_{1}}$ for all $t>0$.  $\square$

In our case $M_{b}$ drops to zero when $t=c^{{-1/2}}$. Therefore, for $b(x)=((x^{2}-m)^{2}+c)^{-1/2}$ we have, for $p>3/2$: 
\begin{multline} \label{norm for bdg}
||b||_{p,w}\leq c^{{-1/2}}c^{{3\over 4p}}({v_d }max(1, {3 (2m)^{1/2} \over{c}^{1/4}}))^{1/p}
\end{multline}
and in particular:
\begin{eqnarray} \label{norm for bdg p3}
||b||_{3,w}\leq c^{-{1\over 4}}({v_3 })^{1/3}max\Big[1, \big({324m^{2}\over{c}}\big)^{1/12} \Big]
\end{eqnarray}

\subsection{Derivation of the main inequality}
Here we repeat almost exactly the analysis of Cancelier, Levy-Bruhl and Nourrigat, ``remarks on the spectrum of dirac operators'' \cite{cancelier1996remarks}. The main difference between our work and theirs is the presence of the chemical potential, which appears in the estimates such as \eqref{norm for bdg p3 del}.

Assume $||V-V_{\lambda}||\leq |\Delta_{0}|\lambda/4$, then
\begin{eqnarray}&
||(H_{0}+V)u||\leq \Delta_{0}\sqrt{1-\lambda}||u|| \Rightarrow \nn &
||(H_{0}+V_{\lambda})u||\leq \Delta_{0}\sqrt{1-\lambda/2}||u||
\end{eqnarray}
To check this note that:
\begin{eqnarray}&
||(H_{0}+V_{\lambda})u||^{2}\leq \nn & ||(H_{0}+V)u||^{2}+2||(H_{0}+V)u||||(V_{\lambda}-V)u||+\nn & ||(V-V_{\lambda})u||^{2}\leq\nn & (|\Delta_{0}|^{2}(1-\lambda)+|\Delta_{0}|^{2}{\lambda\over 2}\sqrt{1-\lambda}+|\Delta_{0}|^{2}{\lambda^{2}\over 16})||u||^{2}\nn & \leq 
|\Delta_{0}|^{2}(1-\lambda+ {\lambda\over 2}\sqrt{1-\lambda}+{\lambda^{2}\over 16})||u||^{2}\leq \nn & |\Delta_{0}|^{2}[1-{\lambda\over 2}]||u||^{2}
\end{eqnarray}
The last inequality can be checked by differentiation.
Thus we have:
\begin{eqnarray*}&
||(H_{0}+V_{\lambda})u||^{2}- |\Delta_{0}|^{2}[1-{\lambda\over 2}]||u||^{2}\leq 0  \Rightarrow \nn &
\la u| (H_{0}^{2}-|\Delta_{0}|^{2})+[H_{0},V_{\lambda}]_{+}+V_{\lambda}^{2}
+|\Delta_{0}|^{2}{\lambda\over 2})|u\ra\leq 0
\end{eqnarray*}
Noting that for BdG $H_{0}^{2}=(-\nabla^{2}-\mu)^{2}{\bf 1}+|\Delta_{0}|^{2}{\bf 1}$, we get for such $u$s:
\begin{eqnarray}&
\la u|(-\nabla^{2}-\mu)^{2}
+|\Delta_{0}|^{2}{\lambda\over 2}+S_{\lambda}|u\ra\leq 0
\end{eqnarray}
where $S_{\lambda}=[H_{0},V_{\lambda}]_{+}+V_{\lambda}^{2}$.
Therefore we are interested in the number of bound states of $(-\nabla^{2}-\mu)^{2}
+|\Delta_{0}|^{2}{\lambda\over 2}+S_{\lambda}$. These are given by a Briman-Schwinger method.
Observing the Birman Schwinger kernel:
\begin{eqnarray}&
K_{BdG}=\nn &{1\over [(-\nabla^{2}-\mu)^{2}
+|\Delta_{0}|^{2}{\lambda\over 2}]^{1/2}}S_{\lambda}{1\over [(-\nabla^{2}-\mu)^{2}
+|\Delta_{0}|^{2}{\lambda\over 2}]^{1/2}}
\end{eqnarray}
We now note that the $S_{\lambda}$ contains several terms, of the following forms (there are also matrix indices, but let us ignore these, since we are going to add the estimate for each element). 
We also assume $\Delta_{0}$ real for simplicity, also, here the perturbation is actually of the form $-\Delta_{1}(x)\sigma_{x}$, so we appropriately define $\Delta_{\lambda}$ as we defined $V_{\lambda}$

Next we use the Cwikel inequalities \eqref{Cwikel inequality} for $p=3$ and the standard inequalities for singular numbers:
\\ {\it Fan's Inequality}:
\begin{eqnarray}
\mu_{n+m+1}(A+B)\leq \mu_{n+1}(A)+\mu_{m+1}(B)
\end{eqnarray}
and
\begin{eqnarray}&\label{product of singular}
\mu_{n}(AB)\leq ||B||\mu_{n}(A)\nn &
\mu_{n+m+1}(AB)\leq \mu_{n+1}(A)\mu_{m+1}(B)
\end{eqnarray}
to deal with the different terms in $K_{{BdG}}$.

We get estimates for the $n$ th singular number of three different forms (here $b$ is as above in \eqref{b bdg}):
\\
1)  \begin{eqnarray}& A_{1}=\Delta_{0}{1\over [(-\nabla^{2}-\mu)^{2}
+|\Delta_{0}|^{2}{\lambda\over 2}]^{1/2}}\sqrt{\Delta_{\lambda}(x)}\times \nn &\sqrt{\Delta_{\lambda}(x)}{1\over [(-\nabla^{2}-\mu)^{2}
+|\Delta_{0}|^{2}{\lambda\over 2}]^{1/2}}\end{eqnarray} \begin{eqnarray}
s_{n}(A_{1})\leq \Delta_{0}\Big(||\sqrt{\Delta_{\lambda}}||_{3}C_{{3,d}}||b||_{3,w}{1\over (n/2)^{1/3}}\Big)^{2}
\end{eqnarray}
Here we used standard inequalities for the singular values of sums and products of operators like \eqref{product of singular}, to have $s_{n}(AB)\leq s_{[n/2]}(A)s_{[n/2]}(B)$.
\\ 2) 
 \begin{eqnarray}& A_{2}=\Delta_{0}{1\over [(-\nabla^{2}-\mu)^{2}
+|\Delta_{0}|^{2}{\lambda\over 2}]^{1/2}} {\Delta_{\lambda}(x)}\times \nn & {\Delta_{\lambda}(x)}{1\over [(-\nabla^{2}-\mu)^{2}
+|\Delta_{0}|^{2}{\lambda\over 2}]^{1/2}}\end{eqnarray} 
Gives in the same way:
\begin{eqnarray}
s_{n}(A_{2})\leq  \Big(||{\Delta_{\lambda}}||_{3}C_{{3,d}}||b||_{3,w}{1\over (n/2)^{1/3}}\Big)^{2}
\end{eqnarray}
\\ 3) 
 \begin{eqnarray}& A_{3}={1\over [(-\nabla^{2}-\mu)^{2}
+|\Delta_{0}|^{2}{\lambda\over 2}]^{1/2}}(-\nabla^{2}-\mu)\times \nn &\Delta_{\lambda}(x){1\over [(-\nabla^{2}-\mu)^{2}
+|\Delta_{0}|^{2}{\lambda\over 2}]^{1/2}}
\end{eqnarray}
since $||{1\over [(-\nabla^{2}-\mu)^{2}
+|\Delta_{0}|^{2}{\lambda\over 2}]^{1/2}}(-\nabla^{2}-\mu)||<1$, we have:
\begin{eqnarray}
s_{n}(A_{3})\leq ||{\Delta_{\lambda}}||_{3}C_{{3,d}}||b||_{3,w}{1\over (n)^{1/3}}
\end{eqnarray}
Combining these estimates and adding them, (and noting that  $||\sqrt{\Delta_{\lambda}}||_{3}^{2}=||\Delta_{\lambda}||_{3/2}$) we conclude the inequality:
\begin{multline}
s_{n}(K_{BdG})\leq C'\Big[ \Delta_{0}||\Delta_{\lambda}||_{3/2}||b||_{3,w}^{2}{1\over n^{2/3}} +\\ ||{\Delta_{\lambda}}||_{3}^{2} ||b||_{3,w}^{2}{1\over n^{2/3}}+||{\Delta_{\lambda}}||_{3}||b||_{3,w}{1\over n^{1/3}}\Big]
\end{multline}
Now we can get the inequality as follows: we are looking for the largest $N$ for which $s_{N}\geq 1$. Thus we multiply by $N^{2/3}$ the above inequality:
\begin{eqnarray}
N^{2/3}  \leq C'\Big[ \Delta_{0}||\Delta_{\lambda}||_{3/2}||b||_{3,w}^{2}  + ||{\Delta_{\lambda}}||_{3}^{2} ||b||_{3,w}^{2}+\\||{\Delta_{\lambda}}||_{3}||b||_{3,w}{N^{1/3}}\Big]
\end{eqnarray}
Thus if $N_{m}$ satisfies:
\begin{eqnarray}&
N^{2/3}_{m}= C'\Big[ \Delta_{0}||\Delta_{\lambda}||_{3/2}||b||_{3,w}^{2}  + \nn & ||{\Delta_{\lambda}}||_{3}^{2} ||b||_{3,w}^{2}+||{\Delta_{\lambda}}||_{3}||b||_{3,w}{N^{1/3}}_{m}\Big]
\end{eqnarray}
Number of bound states must have $N\leq N_{m}$. 
Let us bound $N_{m}$ from above. Writing $N_{m}^{2/3}=A+B N_{m}^{1/3}\Rightarrow N_{m}^{1/3}={B+\sqrt{B^{2}+4A}\over 2}$. Since $\sqrt{B^{2}+4A}\leq B+2\sqrt{A}$, we have: $N_{m}^{1/3}\leq B+\sqrt{A}$. Furthermore, using the generalized mean inequality  $(|a|+|b|+|c|)^{k}\leq 3^{k-1}(|a|^{k}+|b|^{k}+|c|^{k})$ we have:
\begin{eqnarray}&
N_{m}\leq C'\Big[ (\Delta_{0}||\Delta_{\lambda}||_{3/2}||b||_{3,w}^{2})^{3/2}  + \nn & (||{\Delta_{\lambda}}||_{3}^{2} ||b||_{3,w}^{2})^{3/2}+(||{\Delta_{\lambda}}||_{3}||b||_{3,w})^{3}\Big]\leq \nn & C''||b||_{3,w}^{3}\Big[ \Delta_{0}^{3/2}||\Delta_{\lambda}||_{3/2}^{3/2}  + ||{\Delta_{\lambda}}||_{3}^{3}\Big]
\end{eqnarray}
In the BdG context we have $m=\mu$ and $c={\lambda\over 2}|\Delta_{0}|^{2}$ in \eqref{b bdg}. There are two cases:

1) $\mu>\frac{\sqrt{\Delta _0^2-E^2}}{18 \sqrt{2}}$: We have we have
\begin{eqnarray} \label{norm for bdg p3 del}
||b||_{3,w}\leq 2({v_3 })^{1/3}\big({9\mu\over 2{\lambda^{4}}|\Delta_{0}|^{4}}\big)^{1/6}\end{eqnarray}

Using \eqref{norm for bdg p3 del} we have:
\begin{eqnarray} 
||b||_{3,w}^{3}\leq 8{v_3 }\big({9\mu\over 2{\lambda^{4}}|\Delta_{0}|^{4}}\big)^{1/2}
\end{eqnarray}
So finally:
\begin{eqnarray}&
N_{m}\leq  C{\mu^{1/2}\over {\lambda^{2}}|\Delta_{0}|^{2}}\Big[ \Delta_{0}^{3/2}||\Delta_{\lambda}||_{3/2}^{3/2}  + ||{\Delta_{\lambda}}||_{3}^{3}\Big]
\end{eqnarray}
for some $C$.

We can now write this explicitly as an integral:
\begin{eqnarray*}
N_{m}\leq C{\mu^{1/2}\over {\lambda^{2}}|\Delta_{0}|^{2}} \int_{|\delta\Delta(x)|\geq {|\Delta_{0}|\lambda\over 4}} \D x \Big[\Delta_{0}^{3/2}(\delta\Delta_{x})^{3/2}+{(\delta\Delta_{x})^{3}}\Big]
\end{eqnarray*}
The optimal constants in \eqref{Cwikel inequality} are not known in general, but the  various upper-bounds for them are all of order one.
Finally, relating $\lambda$ and $E$ we get \eqref{CLR_for_BdG}. 

2) Similarly, for $\mu <\frac{\sqrt{\Delta _0^2-E^2}}{18 \sqrt{2}}$ and using 
\eqref{norm for bdg p3} we get \eqref{CLR_for_BdG case2}

\bibliographystyle{apsrev}
\bibliography{/Users/iklich/dropbox/Work/KlichBib.bib}
\end{document}